\documentclass[12pt,a4paper]{article}
\usepackage[utf8]{inputenc}
\usepackage{amsmath}
\usepackage{amsfonts}
\usepackage{amssymb}
\usepackage{graphicx}

\title{A threshold model of financial markets}
\author{Paweł Sieczka\footnote{e-mail: psieczka@if.pw.edu.pl}\hspace{0.1cm} and Janusz A. Hołyst}

\begin{document}
\maketitle
\begin{center}
{\it Faculty of Physics, Center of Excellence for Complex Systems Research, Warsaw University of Technology, Koszykowa 75, \\
PL-00-662 Warsaw, Poland}                         \end{center}

\begin{abstract}
We proposed a model of interacting market agents based on the generalized Ising spin model. The agents can take three actions: "buy,"  "sell," or "stay inactive." We defined a price evolution in terms of the system magnetization. The model reproduces main stylized facts of real markets such as: fat-tailed distribution of returns and volatility clustering.
\end{abstract}

\begin{small}\textbf{PACS.} 89.65.Gh                       \end{small}                     

\section{Introduction}
Financial market modelling lies in the field of  interests of both theoreticians and practitioners. Among those who develop these models are also econophysicisits. A view on a financial market as a complex system of investors similar to complex physical systems appeared to be very fruitful, since it allowed to reproduce many characteristic features of such a market \cite{ Bouchaud1, Lux, Stauffer1, Bouchaud2}.

A special group constitute approaches based on the Ising model \cite{Bornholdt1, Kaizoji, Krawiecki, Sornette} or its generalization to a three-state model \cite{Iori, Takaishi, Sato}. They identify agents with spin variables which can take specific values depending on agents' decisions. In much the same way as Ising spins, agents interact with each other, which leads to a herding behavior and, as a consequence, bubbles or crashes. An important target for such models is a reproduction of real market stylized facts such as fat-tailed returns distribution, clustered volatility, or long range correlations of absolute returns \cite{Stanley1}.

In this paper we proposed a simple model of financial markets, based on the Granovetter threshold model of collective behavior,  which is in a good agreement with the above facts.

\section{Description of the model}
An inspiration for our work was the Bornholdt model \cite{Bornholdt2}, corresponding to the Ising spin model with an additional {\it minority term}, where agents act under the influence of their neighbors (Ising part) and a global magnetization. The local field for i-th spin in the Bornholdt model is defined as:
\begin{equation}
h_i=\sum_{j=1}^NJ_{ij}s_j-\alpha s_i\Big |\frac{1}{N}\sum_{j=1}^Ns_j\Big |,
\end{equation}
with a global constant $\alpha>0$.  The model interprets the price of an asset in terms of the magnetization, which enables the authors to reproduce some stylized facts of financial markets.

Here we developed a generalization of the Ising spin model that also uses the absolute value of the magnetization as a factor controlling the dynamics.
 
Let us consider a model of N interacting market agents where each agent  takes one of three actions: "buy," "sell," or "stay inactive." Such an agent can be represented by a three-state spin variable $s_i(t)$ taking values $+1$ when the agent is buying, $-1$ when the agent is selling, and $0$ in the remaining case.  

The agents interact with each other according to an interaction matrix $J_{ij}$. In our model we assumed the interaction matrix corresponding to a 2-dimensional squared lattice in which each agent interacts only with its four nearest neighbors with an equal strength $J$. The interaction strength can be {\it ferromagnetic} ($J>0$) when investors try to act like their neighbors or {\it antiferromagnetic} ($J<0$) when they try to play against their neighbors. We chose the ferromagnetic case to introduce the herding behavior.

At each time step $t$ the agent $i$ takes its value according to the following formula,
\begin{equation} \label{evolution}
s_i(t)={\rm sign}_{\lambda|M(t-1)|}\Big[\sum_{j=1}^N J_{ij}s_j(t-1)+\sigma \eta_i(t)\Big],
\end{equation} 
where ${\rm sign}_q$ is a threshold signum function,
\begin{equation}
{\rm sign_q(x)}=\left \{ \begin{array}{ll}
 1&{\rm if } \;x>q,   \\ 
 0& {\rm if }\; -q<x<q,\\
-1&  {\rm if }\; x<-q.
\end{array}\right.
\end{equation}
The function $\eta_i(t)$ has random values from the Gaussian distribution with $0$ mean and variance equal to $1$. The term $\sigma\eta_i(t)$ simulates an individual erratic opinion of the i-th investor and the parameter $\sigma$ defines strength of individual opinions.

We define the magnetization of the network,
\begin{equation}
M(t)=\frac{1}{N}\sum_{i=1}^N s_i(t),
\end{equation}
which together with a constant $\lambda$ forms a threshold parameter $q$ of the sign function. Let us notice that for $\lambda=0$ our model is identical with the 2-valued Ising spin model.

We simplified the procedure of price calculation presented in \cite{Bornholdt2} by omitting the influence of {\it chartists} and {\it fundamentalists} in the population of investors and redefining the price of an asset as:
\begin{equation}\label{price}
P(t)=P_0(t){\rm e}^{M(t)},
\end{equation}
where $P_0(t)$ is, in agreement with the efficient market hypothesis \cite{Fama}, a geometric Brownian walk corresponding to fundamental price changes. Let us put $P_0(t)=P_0$ constant for more simplicity. We got therefore a logarithmic rate of return:
\begin{equation}
r(t)=M(t)-M(t-1).
\end{equation}

According to (2), each agent is under the influence of three factors. The first is an imitation of their neighbors, associated with the matrix $J_{ij}$ which is responsible for the herding behavior. The second factor is an individual opinion of the agent provided by the term $\sigma\eta_i(t)$. So far, it is the standard Ising model. Yet, there is one more factor, $\lambda|M(t-1)|$,  which plays the role of a threshold parameter. Only those agents that are able to exceed the threshold are allowed to trade. The value of the threshold depends on the absolute magnetization. According to  (5), the magnetization measures a deviation from the fundamental value. So, when it is large, the agents are afraid of trading, unless they have a strong support from the neighbors or from their private opinions.

\section{Results of the simulation}
Using a square $32\times 32$ lattice of agents with periodic boundary conditions we computed the history of the magnetization, and thus the evolution of the stock price.  The agents were allowed to interact only with their nearest neighbors. 

The simulation was started with a random configuration of spins. At each time step a randomly chosen spin was updated according to the evolution equation (2) and this was repeated $N$ times. The first $5000$ time steps were ignored as a period of system thermalization.

We observed that for proper parameters price returns show volatility clustering (fig. 1) and fat tails (fig. 2).

\begin{figure}
\begin{center}
\includegraphics[scale=0.4]{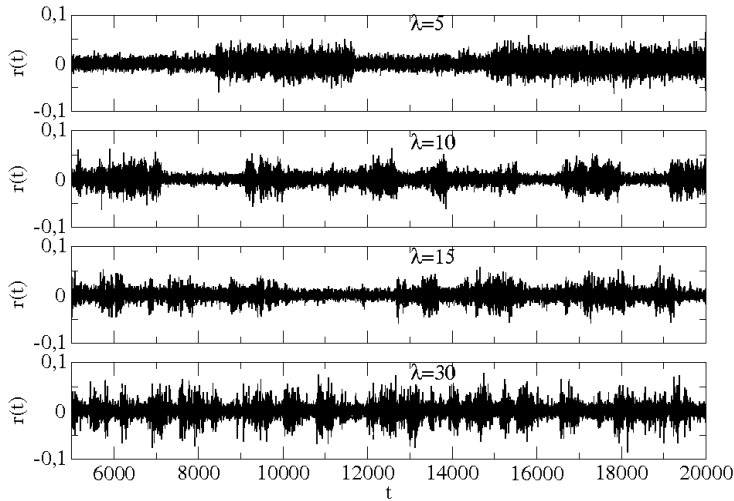}
\caption{Returns $r(t)$ in time $t$ for parameters $J=1$, $\sigma=1$ and different $\lambda$.}
\label{rys1}
\end{center}
\end{figure}
\begin{figure}
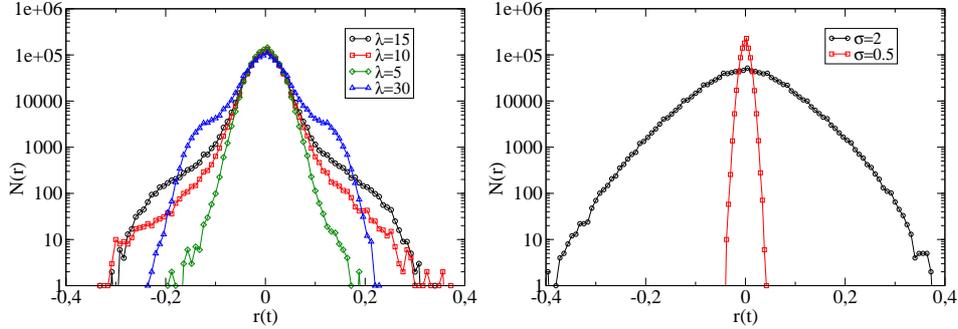

\begin{center}
\begin{tabular}{cc}
\includegraphics[scale=0.25]{rys2a.eps}
\includegraphics[scale=0.25]{rys2b.eps}
\end{tabular}
\caption{(Color online) Histogram of returns $r_{16}(t)$ for parameters $J=1$, $\sigma=1$ and different $\lambda$ (left picture), and for parameters $J=1$, $\lambda=10$, $\sigma=2$, and $\sigma=0.5$ (right picture).}
\end{center}
\end{figure}

\begin{figure}
\centering
 \includegraphics[bb=251 50 554 482, angle=-90, scale=0.8]{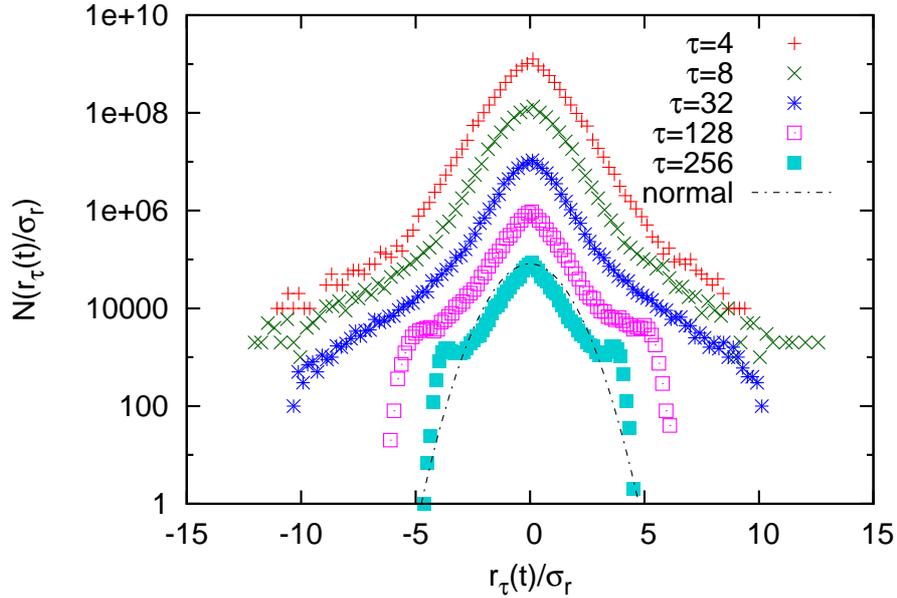}
 \caption{(Color online) Histogram of returns $r_\tau(t)$ for parameters $J=1$, $\sigma=1$, $\lambda=10$, and different time lag $\tau$. The returns were rescalled by the corresponding standard deviation $\sigma_r$. The distributions were shifted upward with the factor 10, 100, 1000, ...  The normal distribution was plotted with a dashed line.}                    
\end{figure}

For real prices, fat tails of returns  distribution ($r_\tau(t)=\log(P(t)/P(t-\tau))$) are getting slimmer with rising time lag $\tau$. Finally, the distribution became normal for sufficiently large $\tau$. In Figure 3 we presented the distribution $r_\tau(t)$ for different $\tau$. The tails are slimming down with rising $\tau$. The distribution has Gaussian tails although its shape is slightly different.

Volatility clustering can be quantitatively shown using the autocorrelation function of a time series $x(t)$:
\begin{equation}
C_x(\tau)=\frac{\langle x(t) x(t-\tau)\rangle - \langle x(t)\rangle^2}{\sigma^2_x},
\end{equation}
where $\sigma_x$ is the variance of $x(t)$, and $\langle...\rangle$ means the average over $t$.  We computed the autocorrelation function of returns $C_r(t)$ (fig. 4) and the absolute value of returns $C_{|r|}(t)$ (fig. 5).   The autocorrelation of returns decays very fast, which is consistent with observations of real markets. It has also the same shape as the distribution presented in [10]. The autocorrelation function of the absolute returns is a slowly decaying exponential similar to the results of [13] and [10]. 

\begin{figure}
\begin{center}
\includegraphics[scale=0.4]{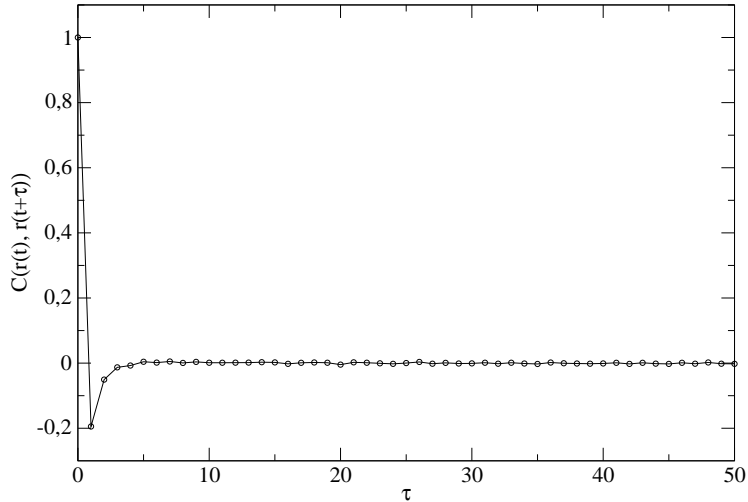}
\caption{Autocorrelation function of return $C_r(\tau)$, obtained for parameters $J=1$, $\sigma=1$, $\lambda=15$.}
\end{center}
\end{figure}

\begin{figure}
\begin{center}
\includegraphics[scale=0.5]{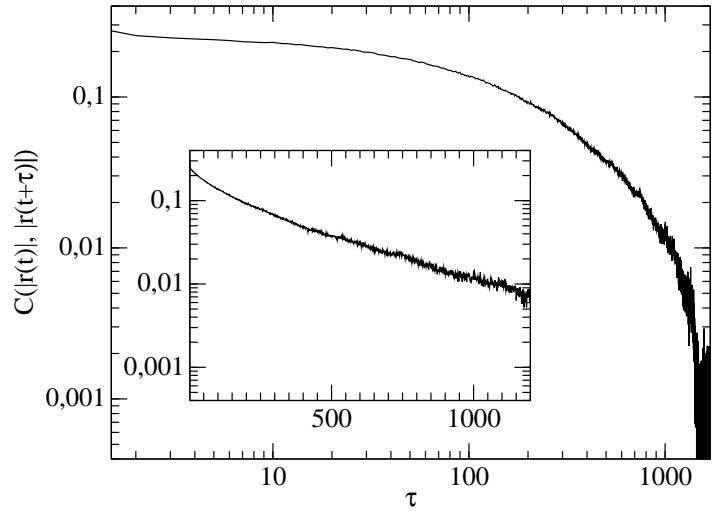}
\caption{Autocorrelation function of absolute return $C_{|r|}(t)$, obtained for parameters $J=1$, $\sigma=1$, $\lambda=15$ in log-log scale, and in semi-log scale (inset). }
\end{center}
\end{figure}

\section{Conclusions}
We proposed a simple model of financial markets based on the Granovetter model. We introduced a threshold controlled by a magnetization of the system. This makes an agent take an action only if its confidence is strong enough to overwhelm the threshold.  

We defined a logarithmic rate of return as a change of the magnetization. The model with such a defined price reproduces main stylized facts of financial markets meaning a fat-tailed distribution of returns, volatility clustering, very fast decaying autocorrelation of returns and much slower decay of autocorrelation of absolute returns. 

It has been observed for real markets that the distribution of returns $r_\tau(t)$ becomes Gaussian for large $\tau$.  The distribution of returns generated by our model approaches the Gaussian distribution for rising $\tau$, however its shape differs from the normal distribution. The autocorrelation function of the absolute returns is a slowly decaying exponential function, while the empirical study of the markets reveals a power-law behavior. Both these issues can be considered as weaker points of our model.

\section*{Acknowledgments}
We would like to dedicate this paper to Prof. Marcel Ausloos and to Prof. Dietrich
Stauffer on the occasion of their 65th birthdays and  to thank them for their inspiring
contributions to econophysics and sociophysics research. \\

The work has been supported by the Project STOCHDYN by European Science Foundation and by Polish Ministry of Science and Higher Education (Grant ESF/275/2006).

\end{document}